# Metis: Multi-Agent Based Crisis Simulation System

George Sidiropoulos[†]
Department of Computer Science,
International Hellenic University
Kavala, Greece
georsidi@teiemt.gr

Chairi Kiourt
Athena-Research & Innovation
Center in Information
Communication and Knowledge
Technologies
Xanthi, Greece
chairiq@athenarc.gr

Lefteris Moussiades
Department of Computer Science,
International Hellenic University
Kavala, Greece
lmous@cs.ihu.gr

## ABSTRACT

With the advent of the computational technologies (Graphics Processing Units - GPUs) and Machine Learning, the research domain of crowd simulation for crisis management has flourished. Along with the new techniques and methodologies that have been proposed all those years, aiming to increase the realism of crowd simulation, several crisis simulation systems/tools have been developed, but most of them focus on special cases without providing users the ability to adapt them based on their needs. Towards these directions, in this paper, we introduce a novel multi-agent-based crisis simulation system for indoor cases. The main advantage of the system is its ease of use feature, focusing on non-expert users (users with little to no programming skills) that can exploit its capabilities a, adapt the entire environment based on their needs (case studies) and set up building evacuation planning experiments with some of the most popular Reinforcement Learning algorithms. Simply put, the system's features focus on dynamic environment design and crisis management, interconnection with popular Reinforcement Learning libraries, agents with different characteristics (behaviors), fire propagation parameterization, realistic physics based on a popular game engine, GPU-accelerated agents training and simulation end conditions. A case study exploiting a popular reinforcement learning algorithm, for training of the agents, presents the dynamics and the capabilities of the proposed systems and the paper is concluded with the highlights of the system and some future directions.

## 1 Introduction

With the advancements of the recent years in computing capabilities, Artificial Intelligence and web technologies the research domain of Crowd Simulation (CS) has gained more and more interest. This field has grown a lot the last decade (and keeps growing) and as a consequence, there are more and more techniques and methods proposed. For example, crowd behaviors simulation [20], emotion contagion management, collision avoidance for pedestrians, accurate decision models etc. are some of the most popular subjects studied as part of the CS domain. Towards these directions, the application of Machine Learning (ML) and especially Deep Learning (DL) approaches have increased and have also been applied in many case studies, with Reinforcement Learning (RL) being an important leader in this studies and closely corelated with CS [17,18,25].

The domain of crowd management and analysis had seen interest as early as 1958 [11], resulting more and more positive social and scientific impact and being continually studied until now [5,13,32]. Most of these studies, focus on developing a level-of-service concepts, designing elements of pedestrian facilities or planning guidelines [12]. Although the goals have remained the same, the demand and simulation scale has increased drastically. Nowadays, the complexity of planning correct emergency evacuations of large and small-scale buildings or building blocks has increased, requiring extensive and accurate planning focusing on different architecture styles, appearances, functionalities and visitor behaviors [27]. All those features (and many other), have become an important aspect of designing a building for efficient evacuation planning, which are important factors in simulations systems for crisis scenarios, for example an evacuation of a building due to fire or earthquake. This type of scenarios aim to improve the procedure of risk assessments, emergency plans and the evacuation itself. Also, they are usually tackled by crisis management preparation procedures, which include mock crisis scenarios (e.g. fire drills or "mock evacuations"). Unfortunately, these types of procedures in many cases fail to prepare humans and are often ignored [7]. Thus, results obtained from those preparation projects cannot be used to design accurate policies. For this reason, simulations systems can be used as an additional method of evaluating a security policy of indoor or outdoor facilities. Simulations can take into account the impact of different environmental, emotional and informational conditions [30], but in most cases the simulation tools have been designed with specific facilities for specific cases

The research domain of Crowd Simulation for Crisis Management (CSCM) has experienced an increasing interest the past years. Crowd simulation is the process of simulating how a number of entities (commonly large) move inside a virtual scene with a specific setting [29]. Crisis simulations are systems that include entities with more roles and responsibilities, on top of the existing techniques and algorithms required for the physical and even psychological simulation of those same entities. Moreover, the setting of the simulated scenario varies a lot, from film production and military simulation to urban planning, which all require high realism concerning the movements of those entities, their grouping and their behaviors in general.

The most suitable approach of crowd and crisis simulation systems is the simulation of multiple individual entities [9]. Systems that follow this approach are called Multi-Agent Systems (MAS) and are consisted of multiple agents (entities to be simulated) and their environment (the setting in which they exist and can interact with each other) in which they may cooperate or compete towards specific tasks/goals [33]. Based on the agents' interactions and their perception they perform actions to achieve their goal. Their structure makes them befitting for crowd and crisis simulation research.

In this paper, a novel crisis simulation system is introduced, focusing towards the creation of a prototype system, that takes advantage of the plethora of simulation and performance enhancing capabilities of a well-known game engine. The system's key features are:

1. Ease of use: users with little to no programming skills/experience can setup a crisis scenario and simulate it through a user-friendly Graphical User Interface (GUI).
2. Dynamic environment design: a feature that allows the users to create their own building and environment based on their needs and case studies.
3. Interconnection with popular Reinforcement Learning libraries: allowing researchers to exploit popular RL algorithms for the training of the agents or to try their own algorithms.
4. Dynamic crisis management: allowing the user to model a specific structured pipeline of a crisis, for example two fires starting from different places.
5. GPU-accelerated agents' training and simulation, for the support of large multi-agent systems.
6. Simulation end options, which allows the user to specify when a simulation will automatically end.

Additionally, we tested the new introduced system with a state-of-the-art Deep Reinforcement Learning algorithm (DRL), resulting high accuracy in training and quite well evacuations of agents in an indoor environment. It should be highlighted that GPU accelerated training of the DRL was much faster than the CPU based training approaches. which boosted the hyperparameters tuning process (time consuming process) of the implemented case study.

Crisis simulation systems have several social and scientific impacts focusing, mainly, in the development of the civilization, by helping humans design safe buildings that gather many visits throughout the day. Additionally, such systems help humans to be prepared for various crisis situations (e.g. evacuation planning in indoor fire cases) by exploiting Artificial Intelligence technologies that simulate human behaviors. In addition to those, this kind of systems, provide persons the ability to design their own environments based on their need and monitor/see how these kinds of scenarios are unfolded.

The rest of the paper is organized as follows: Section 2 briefly presents some crisis simulation platforms and the research that has been done on CSCM, Section 3 introduces the prototype of the new introduced crisis simulation system, followed by Section 4 presenting a case study. Lastly, Section 5 concludes the paper by highlighting some key features of the systems and presents some future work towards the enhancement of the prototype system.

## 2 Related works

As mentioned before, nowadays there has been a plethora of systems developed for the simulation of different crisis scenarios. For example, Becker-Asano et al. presented a multi-agent system focused on first-persona perception and signs, taking dynamically changing occlusions into account [2]. The implementation was done using Unity game engine[1], while also making it possible for participants to be tested in the same virtual airport terminal, with the combination of a head-mounted display "Oculus Rift". Simonov et al. proposed a system for building composite behavior structures for large number of agents [27]. Their system was based on a decision-making algorithm, implemented in Unreal Engine 4[2]. The path finding system exploited the Menge simulation with plugins and the system also included animation support, dynamic models, a visualization module and utility-based strategic level algorithms.

ESCAPES, a multi-agent evacuation simulation system, presented in 2011, which incorporated different agent types with emotional, informational and behavioral interaction [30]. The agent types include individual travelers, families, authority and security agents. Additionally, the system incorporated information spreading to agents, emotional interaction and contagion and the

---

[1] A game engine developed by Unity Technologies (https://unity.com)

[2] A game engine developed by Epic Games. (https://www.unrealengine.com)

Social Comparison Theory [8]. Evakuierungsassistent (translated as Evacuation Assistant) is another simulation system focused on the simulation of evacuation of mass events (e.g. football stadiums), incorporating realistic methods for real-time simulation [31]. The system is agent-based and exploits Cellular Automata (CA) methods and Generalized Centrifugal Force Models [6].

In 2013, De Oliveira Carneiro et al. presented a simulation system to study the crowd's behavior while evacuating a soccer stadium [21]. The system exploits the use of 2D CA defined over multiple grids that represented different levels (state spaces) of simulated environment. The system has the ability to simulate environments with complex structures composed of multiple floors. Sharm et al. [4], proposed the first fire evacuation environment based on the OpenAI gym[3] [26]. Moreover, they proposed a new approach that entails pretraining an agent based on a Deep Q-Network (DQN) algorithm [19] focusing in the discovery of the shortest path to the exit. A very popular platform that adapts to large-scale and complex models is the GAMA platform [28]. It has its own agent-oriented modeling language called Gama Modeling Language (GAML) that follows the object-oriented paradigm. Additionally, the models include spatial components used to represent their 3D representation in the environment. Furthermore, another key feature of the platform is the agent's architecture is based on the Belief Desire Intention (BDI) method [3], that proposes a straightforward formalization of the human reasoning through intuitive concepts. It also supports multi-threaded simulations and running multiple simulations at the same time. Lastly, iCrowd [16] is an agent-based behavior modeling and crowd simulation system that has many different applications, from crowd simulation in crisis evacuations to social behavior and urban/maritime traffic simulation. It makes use of modern, multithreaded and data-oriented approaches that provides architecture extensibility. The system supports studies based on human movements (collision avoidance and path planning) and agent-based behavior modelling.

Several literature reviews and surveys have been published in the last years, presenting advancements and important observations regarding the direction and parts that require focus in the domain of CSCM. Some of the most important stages when developing a simulation tool can be derived by reading some of those reviews. For example, S. Abar's et al. [1] reviewed the literature for quickly assessing the ease of use of a simulation tool. Some of the comparison criteria were the tool's coding language/Application Programming Interface (API), model development effort, modelling strength and scalability level. N. Pelechano and A. Malkawi [22] on their review stated that the physical interactions, psychological elements, improved human movement, agent-based approaches and communication between agents are important features. Lastly, J. Xiao et al. [34] focused on the use of hardware accelerators (especially GPUs) for agent-based simulations.

The presented system is developed taking into account observations/highlights from the aforementioned related works. It has numerous advantages compared to the aforementioned systems, with the most important ones being its ease of use and its usefulness as a research tool for future studies. The development and design of the system focuses on its usability and how easily a user, without any programming skills can setup an environment and test it based on its needs. Moreover, the system can be used as a research tool by scientists to test different RL algorithms/models and apply them on a plethora of different crisis scenarios. This is achieved via the interconnection of Metis with popular RL libraries through the game engine. Additionally, by using a game engine to develop such a system, scalability, physical interactions (physics) and exploitation of GPUs is a given, as games take advantage of those things with very realistic results The combination of these advantages, can highlight it as a unique tool for scientific communities and the general public.

## 3 The "Metis" system

In this section we present a prototype version[4] of a novel multi-agent crisis simulation system, developed over the Unity game engine, called Metis[5][6]. The main structure diagram of Metis is shown in Figure 1, which consists of three major layers: Dynamic Environment Development (DED), Scenario Design (SD) and Evacuation Simulation (ES). The first layer (DED) allows the user to design and setup the entire environment and building to be evacuated, dynamically. The second layer (SD), follows the concept of dynamic design of evacuation scenario, giving the user the ability to place pedestrians (various number of agents) in different parts of the building, designate which doors are exits, mark areas of the environment in which the pedestrians will be safe and place fires in different places. The last layer (ES), handles the evacuation process and the modules responsible for the simulation. The ES layer, exploits machine learning models (interconnection with popular RL libraries), includes the management of the spreading of the fires, handles the ending of simulation and gives the user the ability to manage a dynamically changing crisis. Dynamic crisis management is the ability to model the pipeline of a crisis, in our case two fires from different places. In the future, it could be a tsunami followed by an earthquake etc.

---

[3] A toolkit for developing and comparing reinforcement learning algorithms. (https://gym.openai.com)
[4] The assets currently used are from "Standard Assets (for Unity 2017.3)" (https://assetstore.unity.com/packages/essentials/asset-packs/standard-assets-for-unity-2017-3-32351) and the "Snaps *Prototype | Office*" (https://assetstore.unity.com/packages/3d/environments/snaps-prototype-office-137490) Unity packages.

[5] The name comes from Metis, one of the elder Okeanides and the Titan-goddess of good counsel, planning, cunning and wisdom. Counsel, planning and wisdom are also required when a building is designed. https://www.theoi.com/Titan/TitanisMetis.html
[6] https://sites.google.com/view/metissimulationsystem

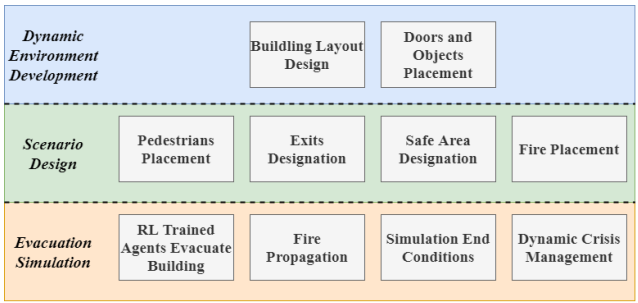

Figure 1: The system framework diagram of Metis.

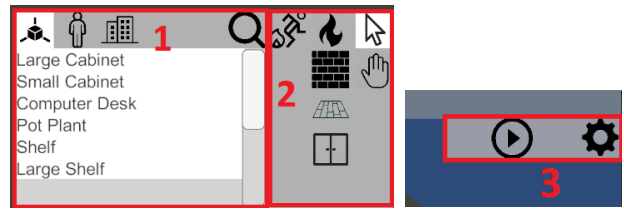

Figure 2: The main User Interface Layers of the Metis system.

A fist-view screenshot of the main component of the system is depicted in Figure 2. The main components of the interface are consisted of 2 User Interface Layers (UIL) and from those, one is also split into two sub-UILs (number 1 and 2 in Figure 2). These UILs focus on the design of the environment, the experiment and in conducting the final evacuation experiments, without the need of expert skills. Also, the main structure of the UILs is based on the framework presented in the Figure 1. Each UIL is consisted of multiple interactive buttons with the following functionalities (left to right, top to bottom):

1. This UIL1 has four main buttons and a scroll view content which includes additional interactive buttons with labels. The three buttons on the top-left ( ) change the category of objects that could be placed in the environment, they appear in the scroll view after a category is selected. The first button ( ) it will show all the static objects that can be placed in the environment, the second ( ) all available types of pedestrians and the last one ( ) all the sample simulation buildings (areas) that can be placed. Sample buildings are buildings created beforehand and provided for the user, with each building including placed objects and having different layout and number of rooms. Clicking on an object in the scroll content will allow the user to place the specific object into the environment. The last button with the magnifier icon ( ) in top-right, allows the user to filter the list of objects through a text field.
2. This UIL2 includes functionalities that can change the mode of the mouse. On the left column the button ( ) allows the user to assign a safe area in the environment. On the middle column the buttons ( ) allow the user to place fires, walls, floors and doors. In the last column the buttons reset the mouse to default ( ) and does nothing and the last button ( ) to grab and place already placed objects.
3. This UIL3 includes buttons regarding the simulation process. The play button starts the simulation process and the gear button shows all the available options regarding the simulation ending conditions.

### 3.1 Dynamic Environment Development

As mentioned above, the DED layer of the system allows the user to design the layout of the building to be evacuated, by placing the building's walls. This layer can be characterized as the layer responsible for the content generation of the environment, commonly used in games [24,35], and allows for the creation of dynamic environments [15] during the environment design process. The walls are placed using the wall placement tool from the UIL1, which places a part of it and can be extended to any direction by dragging the mouse. As a second step, doors can be placed on the walls, allowing pedestrians to move through the rooms. Lastly, a plethora of objects can be placed anywhere inside the building to decorate it and to act as obstacles during the evacuation, which mimic the real-world indoor objects. During any placement procedure, the walls, doors and objects snap to each other so that the placement can be easier. The DED is considered to be a powerful tool, which provides the ability to users to create their own indoor realistic environments based on their needs and their cases. Thus, giving the opportunity to test several different environments without the need of programming skills.

### 3.2 Scenario Design

The SD layer is responsible for designing the scenarios, meaning where the fire will start, how many pedestrians will have to evacuate the building (multi-agent approach) and where their starting positions will be, which exit they will try to reach and where they will be safe. The fires' positions can be chosen with the fire placement tool of the UIL2, which, during the design process, allows the user to specify from which position the fire will start to spread. The fires start spreading when the simulation starts. Different "types" of pedestrians can be placed and each type has different attributes, like speed, size, color, health points etc. giving the ability to simulate different human behaviors. A door is marked as an exit door by right clicking on it. At least one door has to be marked as an exit due to the way the pedestrians were trained. Lastly, safe areas are used to mark a pedestrian as safe from the crisis during the simulation, where they are trying to escape to.

### 3.3 Evacuation Simulation

The ES layer is responsible for all the functions running during the simulation of an evacuation procedure. Starting with the fire propagation, a very simple algorithm is employed. The fire is firstly placed in a point in the building, with a specific maximum area and is represented by a particle emitting object, which damages any object that touches it. Then, when the simulation starts, the area grows periodically and multiple fire objects are created at random places inside the area. Simply put, the propagation of the fire, currently, works with a random speed and direction, while the contact of the fire with the pedestrian is enabled with collisions interfaces. Having control over when the simulation automatically ends is an important feature. The prototype version currently supports end conditions like when all or a specific number of pedestrians are safe/dead.

The pedestrian's evacuation can be done by training the agents with RL algorithms. By exploiting the capabilities of ML-Agents toolkit [14] (Section 4) the agents of the Metis system can be trained with popular RL algorithm such as Proximal Policy Optimization (PPO) [23] and Soft Actor-Critic (SAC) [10]. In addition to that the Metis system can be easily interconnected with popular RL libraries such as RLlib[7] and Baselines[8], also custom python RL algorithms can be developed. A typical RL training is done by creating learning environments in which the agent collects observations and acts based on them. For the training of a general model which can evacuate buildings during a crisis situation, the typical procedure of creating a building environment was followed, setting up doors, designating the exits and placing objects which also acted as obstacles inside the different rooms.

Dynamic crisis management is a part of the simulation that allows the user to manage the crisis currently unfolding. For now, it allows the user to start a fire in a different part of the environment than the initial. This makes the system more effective and allows the user to observe the pedestrians' behaviors while the crisis changes dynamically.

### 3.4 Pedestrian agents training approach

In this section we present and analyze how the pedestrian agents were trained, the features that the agents gather in each step, the actions the agents take to evacuate a building and the environment in which it was trained.

Figure 3 depicts an indoor environment created using the Metis system and then saving it as a single area object for ease of use during the training environment setup and to keep the environment setup for further analysis.

The highlighted areas with light green, inside the building, are the possible areas that an agent could spawn when an episode begun. Initially, all agents spawned inside the room marked as "1" and each one unlocked the next area (light green) one by one. An agent could unlock an area once the mean reward of the RL in the last 20 episodes was equal or higher to 0.925, as consequence of a good training for a specific area. Every time an episode begins, the agent chooses randomly between five possible spawn areas. Those five areas are the most recent areas the agent unlocked. Eventually, when the agent has a mean reward of over 0.925 in average of all areas, it can spawn in any area. This was done so that the agent learned gradually and all the areas were unlocked so that it can generalize correctly (without learning only to escape from the specific point of the building). The red cubes inside the building are dummy fire objects which, when touched, reset the agent's episode and set its reward to -1. The reasoning behind the use of dummy objects instead of the actual fire was to check if the agent, using the aforementioned raycast components, would eventually learn to avoid those fires. While the intense green is considered to be a safe exit for the pedestrians.

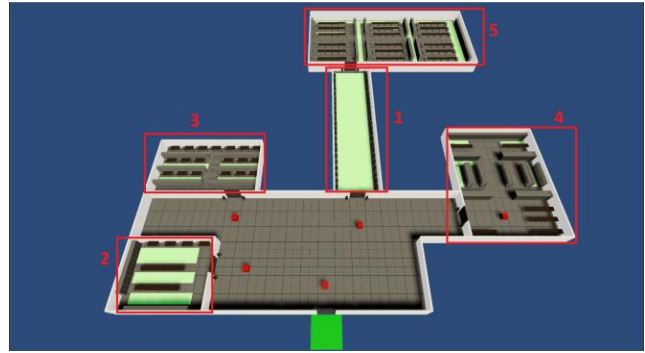

**Figure 3: Building used to train the pedestrian agent.**

The features gathered by each agent during the training/learning were 70 in total, from which 64 were gathered using three "Ray Perception Sensor 3D"[9] components and 6 were calculated manually:

1. The first raycast component detected objects (static objects and fires) and is blocked by walls and doors, it casts 20 rays of 15 length (by default 1:1 meters in Unity), in a 140 degrees arc in front of the agent, responsible for detecting objects that have to be avoided (inside the room the agent currently is).

2. The second raycast component detects doors, safe exit doors and walls, with 20 rays of 25 length in an 80 degrees arc in front of the agent, responsible for detecting, doors and safe exit doors that are close.

3. The third raycast components also detects doors, safe exit doors and walls, with 24 rays of 50 length in a 140 degrees arc in front of the agent, responsible for detecting doors and safe exit doors that are far away.

4. The manually calculated features were:

    a. The normalized x and z values of the safe exit door

    b. The agent's position and

    c. The normalized direction from the agent to the exit door.

---

[7] https://docs.ray.io/en/master/rllib.html
[8] https://github.com/openai/baselines

[9] Rays that are cast into the physics world, and the objects that are hit determine the observation vector that is produced.

During the training, the agent gets -0.4/maxStep reward for each step (action) taken, -0.3/maxStep if collides with something (static objects, walls and closed doors) and small positive rewards depending on its distance from the exit door (Eq. 1). Additionally, when the agent reached the safe area the reward was set to +1.

$$R_{distance} = Distance\left(norm(pos_{exit}), norm(pos_{agent})\right) * 0.3/maxStep \quad (1)$$

Where $norm(pos_{exit})$ the normalized exit's position, $norm(pos_{agent})$ the normalized agent's position, maxStep the maximum number of actions per episode (equal to 10.000 during the training). $Distance()$ calculates the distance between the exit and the agent.

It should be noted that the positions of the objects are normalized according to their relative position inside the building. Additionally, the reasoning behind the choice of the negative rewards is to make the agent reach the exit as soon as possible and to collide with as less objects as possible. Lastly, the reward and episode reset from touching the fire makes the agent avoid the fires and not touch them, by assigning the reward of the episode to -1.

The available actions for each agent are provided through two output branches, each with two possible actions. The first branch is responsible for the agent's horizontal movement (left or right) and the second branch for the vertical movement (backward or forward). With this setup the agent is able to navigate to a safe exit.

Based on these parameters the agents can be trained with many different RL algorithms. By having access to the source code of the Metis system, all these parameters and many others can be adjusted based on the need of the experiment. At this point it should be highlighted that the Metis system will be provided under open source licensing.

## 4 Case Study

For the sake of clarity, a case scenario was setup to present the procedure of setting up a building for a crisis evacuation planning and to also evaluate the quality of the RL based model that was trained in the previous section at escaping a different building in a fire crisis scenario. The creation of the building was done with the UILs and its architecture was quite simple, consisted of three main rooms and one hall. The hall was empty and was connected to the other three rooms (east, north and west rooms). Each room was decorated with many static objects, such as desks, small and large cabinets, small and large shelves and plants. The objects were placed in such way to make the evacuation of the building harder, as it can be seen in the north room in Figure 5. Four doors were placed, three connecting the rooms with the hall and one being the safe exit on the south. The safe exit door was designated as an exit by right clicking on the door. The next step, when setting up a building in the Metis system, is to designate a safe area, which when the pedestrians touch, are considered as being safe. In our case study, we designated a safe are just outside the exit. The following step is to place the pedestrians into the building. We placed a total of 25 pedestrians, scattering them around all the rooms, placing some on difficult areas. Lastly, fires were placed on all rooms, in such way that some pedestrians' paths are partly blocked during the evacuation. Figure 5 shows the layout of our designed building for the case study, along with all the objects, agents and fires placed.

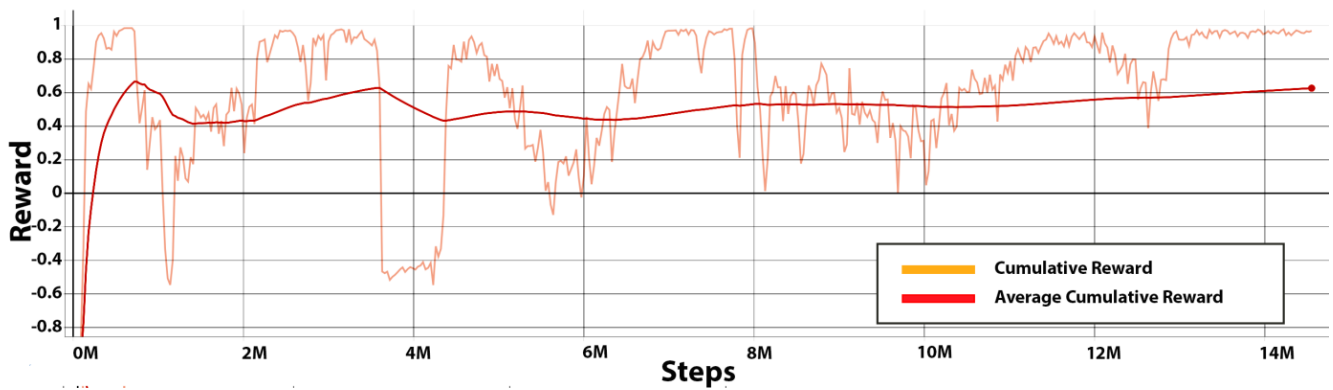

Figure 4: Cumulative reward, an increase during a successful training of agents training.

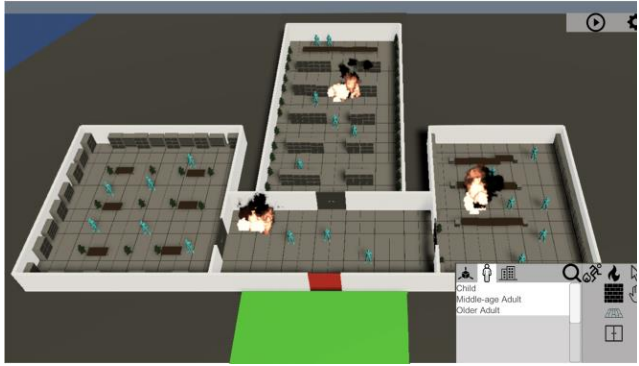

Figure 5: Layout and case study setup.

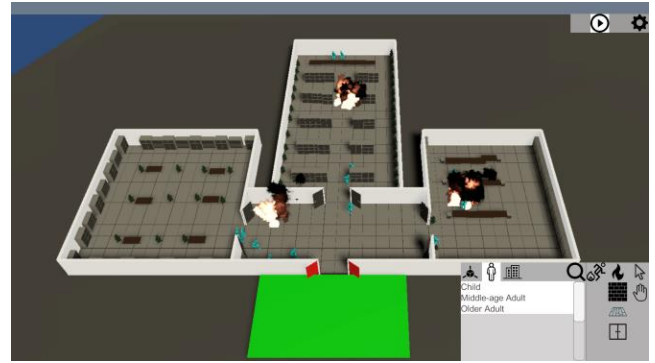

Figure 6: Beginning of the evacuation.

During the training procedure, the environment spawned 60 agents during and each one individually started training. This is a common methodology to speed up the training process. It should be noted that, due to the fact that there were multiple agents in the same environment, they ignored each other, both physically and feature-wise. For the training procedure the PPO algorithm was exploited, which is considered to be one of the most effective RL algorithms for agents' adopting raycast observations.

In our experiment the setup of the chosen PPO algorithm was based on a neural network which approximated the ideal function that mapped the agent's observations to the best action an agent could take in a given state. The neural network set up was, input: 70, hidden layers: 512 and output: 4, with discount factor for future rewards set to $\gamma = 0.995$ and the learning rate set to $\lambda = 0.0003$. Figure 4 depicts the agent's training results. The agent was trained for 14.55 million steps (actions). The reward initially was set to [-1, 1]. The multiple drops in the cumulative reward that can be seen are due to the different "difficulty" areas that were unlocked (new rooms starting points). This, naturally, had as a result to drop the total reward as the environment was different from the previous. Eventually, at the end (where all the areas are unlocked) the cumulative reward reached 0.96.

After the training of the agent was completed, to start the simulation and therefore the building evacuation, the play button was pressed near the top right corner. When this button is pressed, if there is any ongoing crisis (that is, at least one fire has been placed), all the pedestrians will start individually evacuating the building.

Figure 6 shows a snapshot at the beginning of the evacuation after the training, when the pedestrians started running towards the exit. It can be seen that most of the pedestrians find their way towards it immediately. Despite that, some of them can be seen struggling, with some being stuck running into a corner of a room.

When a simulation procedure ends, for any reason, a results window pops up to inform the user about the evacuation and the statistics. Figure 7 shows a snapshot of the info-window, highlighting the results of our case study. The info-window informs the user about the total pedestrians that survived and died at the end of the simulation. In this case study, from the total 25 pedestrians, 17 survived (agents which reached the green safe area) and 8 died (died or didn't evacuate successfully in case of manual end).

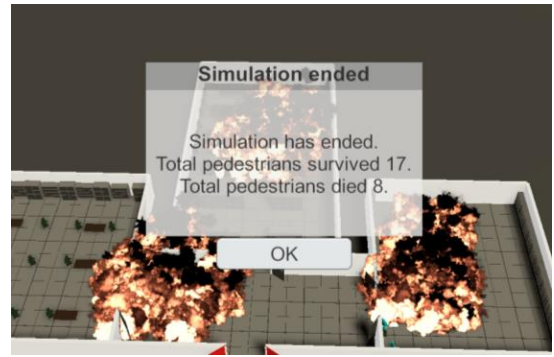

**Figure 7: Window with the simulation's results.**

Note that to demonstrate how easy it is to setup a scenario and to better understand the whole process of creating a building and setting up the evacuation and simulation procedure, we have created a demo video[10].

## 5 Conclusions and future work

Game engines have become more and more popular and have been exploited for many different applications besides their main target, the development of games. In this paper we present a prototype of a novel crisis simulation system called Metis. Metis is developed using a very popular game engine, Unity, and exploits many of its optimizations such as physics, particle effects, cross platform development etc. In addition to that, the Metis system can make use of trending Reinforcement Learning algorithms, to improve the simulation realism and the evacuation planning. Its interconnection with popular RL libraries and its dynamic content (environment) development can establish it as a powerful research tool for basic and applied high-level research. Due to the fact that it is developed over a game engine that supports cross-platform development, it can be considered as a system that can run in multiple operating systems. As mentioned above, the most important key features of the system are its ease of use for scenario design and simulation, the ability to build case study environments dynamically, dynamic management of the crisis situation (multiple crisis situations), exploitation of various RL algorithms and well known libraries, inherent GPU-accelerated agent simulations, agents with various characteristics and behaviors and, lastly, the ability to specify simulation end conditions.

---

[10] https://tinyurl.com/MetisMABCSSDemo

Furthermore, although the system is in Alpha version, the presented experimental results are encouraging and promising. To sum-up, by using the Metis system one can design their own building layout, place a variety of objects, agents and fires, towards the development of personal evacuation plan. Due to its simplicity the Metis system can be used by everyone, even from users without special programming skills. The aforementioned features of Metis focus on a key concept of a dynamic and general system, especially due to the dynamic crisis management. The user can start a crisis at different moments during the simulation, creating unique scenarios and allowing them to observe the pedestrians' reactions.

From the results it is obvious that there is room for improvement. First and foremost, not all pedestrians found their way towards the exit, which means a different training approach with different RL algorithms has to be tested. The best one would be to train the agent for much longer, with a dynamic change of the environment (random or via curriculum learning). This means that the placed fires' positions have to be changed, along with the exit door, the building's layout and the agent's attributes (speed and size), every time the agent finishes an episode. In general, a better fine-tuning of the algorithms could provide more accurate evacuations with fewer losses. Moreover, allowing pedestrians to interact with each other (cooperative learning) will require the agents to be trained in such way that they take into account the number of agents near them or near an exit. In addition to that, future work on the system includes the introduction of other features and functionalities, such as: real time simulation statistics, more explanatory and graphical statistics at the end of a simulation, ability to build multi-level and multiple buildings, more realistic fire propagation, more types of crisis (in addition to fire, such as earthquake, flooding etc.) and allow pedestrians to interact with each other. Moreover, some important considerations towards the future improvement of the system are the incorporation of emotional and psychological features into the agents. This aspect is an important one and has been extensively studied in the literature of CSCM. Lastly, an important feature of a system that aims for longevity and extensibility is to add support for the user (auto guide) to extend the system's functionalities.

## ACKNOWLEDGMENTS

This work is partially supported by the MPhil program "Advanced Technologies in Informatics and Computers", hosted by the Department of Computer Science, International Hellenic University.

## REFERENCES

[1] Sameera Abar, Georgios K. Theodoropoulos, Pierre Lemarinier, and Gregory M.P. O'Hare. 2017. Agent Based Modelling and Simulation tools: A review of the state-of-art software. *Comput. Sci. Rev.* 24, (2017), 13–33. DOI:https://doi.org/10.1016/j.cosrev.2017.03.001

[2] Christian Becker-Asano, Felix Ruzzoli, Christoph Hölscher, and Bernhard Nebel. 2014. A multi-agent system based on unity 4 for virtual perception and wayfinding. *Transp. Res. Procedia* 2, (2014), 452–455. DOI:https://doi.org/10.1016/j.trpro.2014.09.059

[3] Lars Braubach, Alexander Pokahr, and Winfried Lamersdorf. 2005. Jadex: A BDI-Agent System Combining Middleware and Reasoning. *Softw. Agent-Based Appl. Platforms Dev. Kits* (2005), 143–168. DOI:https://doi.org/10.1007/3-7643-7348-2_7

[4] Greg Brockman, Vicki Cheung, Ludwig Pettersson, Jonas Schneider, John Schulman, Jie Tang, and Wojciech Zaremba. 2016. OpenAI Gym. (2016), 1–4. Retrieved from http://arxiv.org/abs/1606.01540

[5] R. L. Carstens and S. L. Ring. 1970. Pedestrian capacities of shelter entrances. *Traffic Eng.* 41, 3 (1970), 38–43.

[6] Mohcine Chraibi, Armin Seyfried, and Andreas Schadschneider. 2010. Generalized centrifugal-force model for pedestrian dynamics. *Phys. Rev. E - Stat. Nonlinear, Soft Matter Phys.* 82, 4 (2010). DOI:https://doi.org/10.1103/PhysRevE.82.046111

[7] R. Fahy and G. Proulx. 1997. Human Behavior In The World Trade Center Evacuation. *Fire Saf. Sci.* 5, (1997), 713–724. DOI:https://doi.org/10.3801/IAFSS.FSS.5-713

[8] Leon Festinger. 1954. A Theory of Social Comparison Processes. *Hum. Relations* 7, 2 (May 1954), 117–140. DOI:https://doi.org/10.1177/001872675400700202

[9] G. Nigel. Gilbert and Klaus G. Troitzsch. 2005. *Simulation for the social scientist*.

[10] Tuomas Haarnoja, Aurick Zhou, Kristian Hartikainen, George Tucker, Sehoon Ha, Jie Tan, Vikash Kumar, Henry Zhu, Abhishek Gupta, Pieter Abbeel, and Sergey Levine. 2018. Soft Actor-Critic Algorithms and Applications. (December 2018). Retrieved from http://arxiv.org/abs/1812.05905

[11] B. D. Hankin and R. A. Wright. 1958. Passenger Flow in Subways. *OR* 9, 2 (June 1958), 81. DOI:https://doi.org/10.2307/3006732

[12] Dirk Helbing, Péter Molnár, Illés J. Farkas, and Kai Bolay. 2001. Self-organizing pedestrian movement. *Environ. Plan. B Plan. Des.* 28, 3 (2001), 361–383. DOI:https://doi.org/10.1068/b2697

[13] L. A. Hoel. 1968. Pedestrian travel rates in central business districts. *Traffic Eng.* (1968), 10–13.

[14] Arthur Juliani, Vincent-Pierre Berges, Esh Vckay, Yuan Gao, Hunter Henry, Marwan Mattar, and Danny Lange. 2018. Unity: A General Platform for Intelligent Agents. (September 2018). Retrieved from http://arxiv.org/abs/1809.02627

[15] Chairi Kiourt, Anestis Koutsoudis, and George Pavlidis. 2016. DynaMus: A fully dynamic 3D virtual museum framework. *J. Cult. Herit.* 22, (November 2016), 984–991. DOI:https://doi.org/10.1016/j.culher.2016.06.007

[16] Vassilios I. Kountouriotis, Manolis Paterakis, and Stelios C. A. Thomopoulos. 2016. iCrowd: agent-based behavior modeling and crowd simulator. 98420Q. DOI:https://doi.org/10.1117/12.2223109

[17] Sheng Yan Lim. 2015. Crowd Behavioural Simulation via Multi-Agent Reinforcement Learning.

[18] Francisco Martinez-Gil, Miguel Lozano, and Fernando Fernández. 2014. Strategies for simulating pedestrian navigation with multiple reinforcement learning agents. *Auton. Agent. Multi. Agent. Syst.* 29, 1 (2014), 98–130. DOI:https://doi.org/10.1007/s10458-014-9252-6

[19] Volodymyr Mnih, Koray Kavukcuoglu, David Silver, Alex Graves, Ioannis Antonoglou, Daan Wierstra, and Martin Riedmiller. 2013. Playing Atari with Deep Reinforcement Learning. (2013), 1–9. Retrieved from http://arxiv.org/abs/1312.5602

[20] Rahul Narain, Abhinav Golas, Sean Curtis, and Ming C. Lin. 2009. Aggregate Dynamics for Dense Crowd Simulation. *ACM Trans. Graph.* 28, 5 (2009), 1–8. DOI:https://doi.org/10.1145/1618452.1618468

[21] Lílian De Oliveira Carneiro, Joaquim Bento Cavalcante-Neto, Creto Augusto Vidal, and Teófilo Bezerra Dutra. 2013. Crowd evacuation using cellular automata: Simulation in a soccer stadium. *Proc. - 2013 15th Symp. Virtual Augment. Reality, SVR 2013* (2013), 240–243.

DOI:https://doi.org/10.1109/SVR.2013.29
[22] Nuria Pelechano and Ali Malkawi. 2008. Evacuation simulation models: Challenges in modeling high rise building evacuation with cellular automata approaches. *Autom. Constr.* 17, 4 (2008), 377–385. DOI:https://doi.org/10.1016/j.autcon.2007.06.005
[23] John Schulman, Filip Wolski, Prafulla Dhariwal, Alec Radford, and Oleg Klimov. 2017. Proximal Policy Optimization Algorithms. (July 2017). Retrieved from http://arxiv.org/abs/1707.06347
[24] Noor Shaker, Julian Togelius, and Mark J. Nelson. 2016. *Procedural Content Generation in Games*. Springer International Publishing, Cham. DOI:https://doi.org/10.1007/978-3-319-42716-4
[25] Jivitesh Sharma, Per-Arne Andersen, Ole-Chrisoffer Granmo, and Morten Goodwin. 2019. Deep Q-Learning with Q-Matrix Transfer Learning for Novel Fire Evacuation Environment. (2019), 1–21. Retrieved from http://arxiv.org/abs/1905.09673
[26] Jivitesh Sharma, Per-Arne Andersen, Ole-Christoffer Granmo, and Morten Goodwin. 2020. Deep Q-Learning With Q-Matrix Transfer Learning for Novel Fire Evacuation Environment. *IEEE Trans. Syst. Man, Cybern. Syst.* (2020). DOI:https://doi.org/10.1109/tsmc.2020.2967936
[27] Andrey Simonov, Aleksandr Lebin, Bogdan Shcherbak, Aleksandr Zagarskikh, and Andrey Karsakov. 2018. Multi-agent crowd simulation on large areas with utility-based behavior models: Sochi Olympic Park Station use case. *Procedia Comput. Sci.* 136, (2018), 453–462. DOI:https://doi.org/10.1016/j.procs.2018.08.266
[28] Patrick Taillandier, Benoit Gaudou, Arnaud Grignard, Quang-Nghi Huynh, Nicolas Marilleau, Philippe Caillou, Damien Philippon, and Alexis Drogoul. 2019. Building, composing and experimenting complex spatial models with the GAMA platform. *Geoinformatica* 23, 2 (April 2019), 299–322. DOI:https://doi.org/10.1007/s10707-018-00339-6
[29] Daniel Thalmann. 2016. Crowd Simulation. In *Encyclopedia of Computer Graphics and Games*. Springer International Publishing, Cham, 1–8. DOI:https://doi.org/10.1007/978-3-319-08234-9_69-1
[30] Jason Tsai, Natalie Fridman, Emma Bowring, Matthew Brown, Shira Epstein, Gal Kaminka, Stacy Marsella, Andrew Ogden, Inbal Rika, Ankur Sheel, Matthew E. Taylor, Xuezhi Wang, Avishay Zilka, and Milind Tambe. 2011. ESCAPES - Evacuation simulation with children, authorities, parents, emotions, and social comparison. *10th Int. Conf. Auton. Agents Multiagent Syst. 2011, AAMAS 2011* 1, (2011), 425–432.
[31] Armel Ulrich Kemloh Wagoum, Mohcine Chraibi, Jonas Mehlich, Armin Seyfried, and Andreas Schadschneider. 2012. Efficient and validated simulation of crowds for an evacuation assistant. *Comput. Animat. Virtual Worlds* 23, 1 (February 2012), 3–15. DOI:https://doi.org/10.1002/cav.1420
[32] Ulrich Weidmann. 1993. *Transporttechnik der Fussgänger*. DOI:https://doi.org/https://doi.org/10.3929/ethz-a-010025751
[33] Michael Wooldridge. 2009. *An Introduction to MultiAgent Systems, 2nd Edition*.
[34] Jiajian Xiao, Philipp Andelfinger, David Eckhoff, Wentong Cai, and Alois Knoll. 2019. A survey on agent-based simulation using hardware accelerators. *ACM Comput. Surv.* 51, 6 (2019). DOI:https://doi.org/10.1145/3291048
[35] Georgios N. Yannakakis and Julian Togelius. 2018. *Artificial Intelligence and Games*. Springer International Publishing, Cham. DOI:https://doi.org/10.1007/978-3-319-63519-4